\documentclass[reprint,aps,longbibliography,superscriptaddress,showpacs,preprintnumbers,amsmath,amssymb,twocolumn,notitlepage]{revtex4-1}

\usepackage{graphicx}
\usepackage{dcolumn}
\usepackage{bm}
\usepackage{color}
\usepackage{ulem}
\usepackage{amssymb}
\usepackage{enumerate}
\usepackage{floatrow}
\usepackage{graphicx}
\usepackage{subfig}
\usepackage[colorlinks,
linkcolor=red,
anchorcolor=black,
citecolor=blue]{hyperref}
\usepackage{tikz}
\usetikzlibrary{shapes,snakes}
\usepackage{ragged2e}
\begin{document}

\title{Intrinsically interacting higher-order topological superconductors}

\author{Hao-Ran Zhang}
\thanks{These authors contributed equally.}
\affiliation{State Key Laboratory of Low Dimensional Quantum Physics and Department of Physics, Tsinghua University, Beijing 100084, China}
\author{Jian-Hao Zhang}
\thanks{These authors contributed equally.}
\affiliation{Department of Physics, The Chinese University of Hong Kong, Shatin, New Territory, Hong Kong, China}
\affiliation{Department of Physics, The Pennsylvania State University, University Park, Pennsylvania 16802, USA}
\author{Zheng-Cheng Gu}
\email{zcgu@phy.cuhk.edu.hk}
\affiliation{Department of Physics, The Chinese University of Hong Kong, Shatin, New Territory, Hong Kong, China}
\author{Rui-Xing Zhang}
\email{ruixing@utk.edu}
\affiliation{Department of Physics and Astronomy, The University of Tennessee, Knoxville, Tennessee 37996, USA}
\affiliation{Department of Materials Science and Engineering, The University of Tennessee, Knoxville, Tennessee 37996, USA}
\affiliation{Institute for Advanced Materials and Manufacturing, The University of Tennessee, Knoxville, Tennessee 37920, USA}
\author{Shuo Yang}
\email{shuoyang@tsinghua.edu.cn}
\affiliation{State Key Laboratory of Low Dimensional Quantum Physics and Department of Physics, Tsinghua University, Beijing 100084, China}
\affiliation{Frontier Science Center for Quantum Information, Beijing 100084, China}
\affiliation{Hefei National Laboratory, Hefei 230088, China}

\date{\today}

\begin{abstract}
We propose a minimal interacting lattice model for two-dimensional class-$D$ higher-order topological superconductors with no free-fermion counterpart.
A Lieb-Schultz-Mattis-type constraint is proposed and applied to guide our lattice model construction. 
Our model exhibits a trivial product ground state in the weakly interacting regime, whereas, increasing electron interactions provoke a novel topological quantum phase transition to a $D_4$-symmetric higher-order topological superconducting state. 
The symmetry-protected Majorana corner modes are numerically confirmed with the matrix-product-state technique.
Our theory paves the way for studying interacting higher-order topology with explicit lattice model constructions.
\end{abstract}

\maketitle
\newcommand{\lra}{\longrightarrow}
\newcommand{\xra}{\xrightarrow}
\newcommand{\ra}{\rightarrow}
\newcommand{\bs}{\boldsymbol}
\newcommand{\ul}{\underline}
\newcommand{\1}{\text{\uppercase\expandafter{\romannumeral1}}}
\newcommand{\2}{\text{\uppercase\expandafter{\romannumeral2}}}
\newcommand{\3}{\text{\uppercase\expandafter{\romannumeral3}}}
\newcommand{\4}{\text{\uppercase\expandafter{\romannumeral4}}}
\newcommand{\5}{\text{\uppercase\expandafter{\romannumeral5}}}
\newcommand{\6}{\text{\uppercase\expandafter{\romannumeral6}}}

\textit{Introduction.} 
The discovery of the quantum Hall effect~\cite{FQHE, Laughlin} raised the curtain on one of the greatest triumphs of condensed-matter physics, the topological phases of matter. 
Different from traditional Landau's paradigm, phases with topological distinctions are characterized by their patterns of long-range entanglement rather than symmetry-breaking orders. 
Moreover, lattice or internal symmetries can help short-range entangled states develop additional topological structures, leading to symmetry-protected topological (SPT) phases \cite{ZCGu2009, chen11a, 230, XieChenScience, cohomology, Senthil_2015, E8, Lu12, invertible2, invertible3, special, general1, general2, Kapustin2014, Kapustin2015, Kapustin2017, Gu-Levin, gauging1, gauging3, dimensionalreduction, gauging2, 2DFSPT, braiding, Ning21a}. 
Recently, SPT phases protected by crystalline symmetries have been under the research spotlight~\cite{TCI, Fu2012, ITCI, reduction, building, correspondence, SET, 230, BCSPT, Jiang2017, Kane2017, slager2013space, Shiozaki_2022, ZDSong2018, defect, realspace, shiozaki2018generalized, rotation, dihedral, LuX, YMLu2018, Hermele2018, Po2020, Huang2020PRR, Huang2021PRR, wallpaper, PEPS, Maissam2020, Maissam2021, Max18, PhysRevX.6.041068, Meng18fLSM, JosephMeng19, BBC, 3Dpoint}, mainly attributed to their capability of supporting an exotic {\it higher-order} topological bulk-boundary correspondence. 
Namely, an $n$th order topological phenomenon takes place when $(d-n)D$ gapless boundary modes ($1<n\leq d$) show up in a $dD$ crystalline SPT phase~\cite{Wang2018, Yan2018, Nori2018, Wangyuxuan2018, Ryu2018, Zhang2019, zhang2019higher, zhang2020mobius, Hsu2018, bultinck2019three, Roy_2020, Roy_2021, Laubscher_2019, Laubscher_2020, Zhang_2022, May-Mann2022,zhang2022bulk,PhysRevB.104.134508,SSPT2022}. 
In particular, a  two-dimensional (2D) higher-order topological superconductor (HOTSC) will, by definition, host non-Abelian Majorana modes that reside around the sample geometric corners, and recent studies have proposed schemes to realize the non-Abelian braiding of Majorana corner modes ~\cite{PhysRevResearch.2.043025, PhysRevB.102.100503}, which offers a new platform to topologically encode quantum information. 

The state-of-the-art development of topological band theory phrased as \textit{symmetry indicators} \cite{indicator1, indicator2, indicator3, indicator5} has boosted our understanding of HOTSCs at the free-fermion level with a plethora of candidate systems having been theoretically proposed. 
Nonetheless, no convincing experimental evidence of HOTSC physics has been reported, thus far. 
Although the symmetry indicator theory only applies to free-fermion systems, the formation of 
HOTSC usually requires an unconventional pairing symmetry, which likely arises from electron interaction effects.
Therefore, a faithful microscopic theory or prediction of HOTSC would also require comprehensive knowledge of relevant topological physics in the strongly interacting regime. 
Along this direction, recent works have established a constructive classification scheme of crystalline fermion SPT phases (e.g., Refs.~\cite{rotation,dihedral,wallpaper,3Dpoint}), which involves novel topological structures that can be interpreted as interacting HOTSC phases with no free-fermion counterpart. 
However, the real-space constructions in the classification are relatively abstract, leaving it unclear whether and how these new and exciting ideas of {\it intrinsically} interacting HOTSCs can be modeled in a concrete lattice system, let alone realized in a laboratory. This motivates us to design explicit lattice models to realize these exotic phases. With the explicit lattice model, hopefully it will become more possible to explore this intrinsically interacting HOTSC physics in experiments and facilitate the development of Majorana-based quantum computation.

In this Letter, we construct a minimal lattice model of 2D HOTSC that cannot be realized in any free-fermion systems. 
The SPT nature of our model is supported by a Lieb-Schultz-Mattis-type (LSM-type) constraint, the original version of which forbids a unique gapped symmetric ground state in a one-dimensional (1D) spin-$1/2$ chain with translation symmetry and $SO(3)$ on-site symmetry \cite{Lieb_1961}, and can be generalized to higher dimensions and other internal symmetries \cite{Oshikawa_2000,Hastings_2004,Hastings_2005}. 
The LSM-type constraint can be viewed as the certain bulk-boundary correspondence of weak SPT phase in one higher dimension, or, alternatively, can be interpreted as a mixed anomaly between translation symmetry and internal symmetry~\cite{PhysRevX.6.041068, Meng18fLSM}.
The LSM-type theorem reveals that the microscopic structure of a lattice system can impose strong constraints on its low-energy behavior.
It, therefore, serves as a useful criterion to help us exclude lattice constructions that are SPT impossible (i.e., forbid a unique gapped symmetric ground state) and further guides us to the ``correct" lattice models with desired topological properties.
Building upon this starting point, we begin with a free-fermion Hamiltonian in a designed lattice and add proper $D_4$-symmetric interactions to it.
In the absence of interactions, our model exhibits a trivial gapped phase, which persists when moderate interactions are turned on.
Further increasing the interaction strength triggers a novel topological quantum phase transition between the trivial phase and the HOTSC phase, the critical behavior of which has been carefully studied here. 
We also confirm the signature Majorana corner modes of the HOTSC phase numerically by placing our model on an open-boundary two-leg ladder geometry with the help of the matrix-product-state (MPS) technique.
Our lattice model is an important step towards bridging the gap between the formal classification theory and the materialization of strongly interacting higher-order topological physics.

\textit{LSM-type constraint.} 
The LSM-type constraint was originally proposed in 1D spin systems with translation symmetry and later generalized to 2D translationally invariant systems with various internal symmetries\cite{Lieb_1961,Oshikawa_2000,Hastings_2004,Hastings_2005}. 
It is natural to generalize the LSM-type constraint to systems with general crystalline symmetry.
However, there is a significant difference between crystalline symmetry and internal symmetry. 
In systems with crystalline symmetry, the effective on-site symmetry, often known as a {\it site symmetry group}~\cite{bradlyn2017topological}, can vary across different spatial locations. 
One may simplify this situation by placing physical degrees of freedom only on the maximal Wyckoff positions of the lattice.
This simplification is supported by the fact that physical degrees of freedom if not being maximally Wyckoff positioned can always be smoothly deformed to these maximal Wyckoff positions symmetrically through a lattice homotopy~\cite{PhysRevLett.119.127202}.
As a result, the LSM-type constraint for crystalline-symmetric lattice systems should be defined for the maximal Wyckoff positions.

The LSM-type constraint for topological crystalline phases in 2D interacting fermionic systems is defined as a gapped nondegenerate ground state requires that the system can be adiabatically connected to a state with an integer multiple of linear representations of the total symmetry group at maximal Wyckoff positions per unit cell \cite{supplementary}. 
In other words, if the physical degrees of freedom within a unit cell can be deformed to a projective representation of the total symmetry group at maximal Wyckoff positions, the ground state either breaks symmetry or has gapless excitations. 
In Supplemental Material, we demonstrate the LSM-type constraint in 2D fermionic systems with $D_2$ symmetry \cite{supplementary}. A lattice system that satisfies the LSM-type constraint forbids a unique gapped symmetric ground state, and, thus, cannot be a candidate for SPT phase.
As a result, the LSM-type constraint immensely reduces the possibilities of assigning physical degrees of freedom on the lattice and greatly simplifies the model construction of the higher-order crystalline topological phase.

\begin{figure}
\centering
\includegraphics[width=1.0\textwidth]{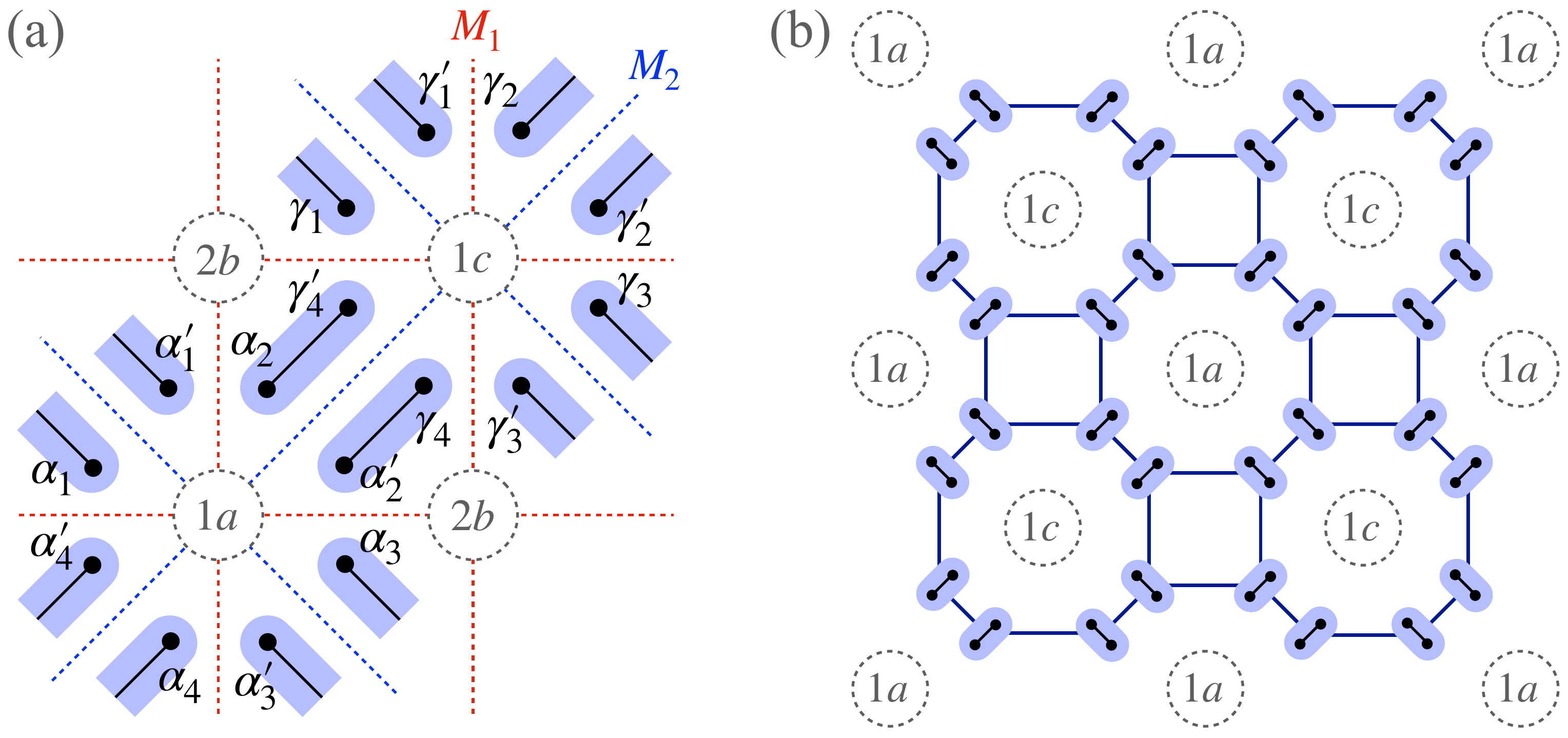}
\caption{
(a) Reflection generators $\bs{M}_{1,2}$ and maximal Wyckoff positions (dashed circles) $a,b,c$ of $D_4$. 
Blue shadows depict atomic sites assigned between each neighboring maximal Wyckoff positions $a$ and $c$, composed of two Majorana zero modes depicted by two solid dots. 
(b) Lattice in the open boundary condition, where $1a/1c$ denotes the maximal Wyckoff position. Atomic sites are placed on a square-octagon lattice depicted by solid lines.
}
\label{FIG1}
\end{figure}

\textit{$D_4$-symmetric HOTSC.} 
Now, we apply the LSM-type constraint to a system with $D_4$-crystalline symmetry. The fourfold dihedral group $D_4$ is the semidirect product of a fourfold rotation group $C_4$ and a reflection group $\mathbb{Z}_2^M$ (i.e., $D_4=C_4\rtimes\mathbb{Z}_2^M$). 
Alternatively, as illustrated in Fig. \ref{FIG1}(a), $D_4$ can be generated by two reflection operators $\bs{M}_1$ and $\bs{M}_2$. 
There are three maximal Wyckoff positions per unit cell, denoted as $\bs{q}_a=(0,0)$, $\bs{q}_{b}=(0,1/2)$, or $(1/2,0)$ and $\bs{q}_c=(1/2,1/2)$. 
Here, the site-symmetry group for both $a$ and $c$ is $D_4$, whereas, that for $b$ is $D_2$. 
To build a $D_4$-symmetric 2D HOTSC with a unique gapped ground state, the LSM-type constraint requires an even number of (pseudo) spin-$1/2$ degrees of freedom at maximal Wyckoff positions per unit cell because a spin-$1/2$ degree of freedom forms a projective representation of a $D_4$ group. 
Note that a pair of spinless fermions, or equivalently, four Majorana fermions, can form a spin-$1/2$ operator. 
Thus, we should place, at least, eight Majorana operators at each occupied maximal Wyckoff position for a gapped SPT state to emerge.

Following that, we consider a lattice model with $16$ Majorana operators per unit cell, eight at each maximal Wyckoff position, $a/c$ which form a linear representation of $\mathbb{Z}_2^f\times D_4$. 
As shown in Fig.~\ref{FIG1}(a), we label these Majorana fermions $\alpha_k$, $\alpha_k'$, $\gamma_k$, and $\gamma_k'$, with $k=1,2,3,4$. To describe a physical superconductor, the Majorana fermions must come in pairs to form complex fermions corresponding to atomic sites (denoted by the blue shadows), which sit around the midpoints between Wyckoff positions $a$ and $c$. 
As shown in Fig. \ref{FIG1}(b), the complex fermions together form a square-octagon lattice.
We construct a $D_4$-symmetric Hamiltonian with intraite Majorana pairs denoted by the black solid lines in Fig.~\ref{FIG1} (a),
\begin{equation}
\begin{aligned}
H_0 &= it \sum_j (
\alpha_{2,j}\gamma^{\prime}_{4,j} +  \alpha_{2,j}^{\prime}\gamma_{4,j} + \alpha_{3,j}\gamma^{\prime}_{1,j-\hat{y}} + \alpha_{3,j}^{\prime}\gamma_{1,j-\hat{y}}\\
+&\alpha_{4,j}\gamma^{\prime}_{2,j-\hat{x}-\hat{y}} + \alpha_{4,j}^{\prime}\gamma_{2,j-\hat{x}-\hat{y}} +  \alpha_{1,j}\gamma^{\prime}_{3,j-\hat{x}}+\alpha_{1,j}^{\prime}\gamma_{3,j-\hat{x}}
).
\end{aligned}
\label{xyterm}
\end{equation}
The additional subscript $j$ of the Majorana operator labels the unit cell. 
One can easily verify that $H_0$ is invariant under $D_4$ action via the following symmetry transformations:
\begin{equation}
\begin{aligned}
& \boldsymbol{M}_1: \begin{array}{l}
\left(\alpha_1, \alpha_2, \alpha_3, \alpha_4\right) \leftrightarrow\left(\alpha_2^{\prime}, \alpha_1^{\prime}, \alpha_4^{\prime}, \alpha_3^{\prime}\right) \\
\left(\gamma_1, \gamma_2, \gamma_3, \gamma_4\right) \leftrightarrow\left(\gamma_2^{\prime}, \gamma_1^{\prime}, \gamma_4^{\prime}, \gamma_3^{\prime}\right)
\end{array} \\
& \boldsymbol{M}_2: \begin{array}{l}
\left(\alpha_1, \alpha_2, \alpha_3, \alpha_4\right) \leftrightarrow\left(\alpha_3^{\prime}, \alpha_2^{\prime}, \alpha_1^{\prime}, \alpha_4^{\prime}\right) \\
\left(\gamma_1, \gamma_2, \gamma_3, \gamma_4\right) \leftrightarrow\left(\gamma_3^{\prime}, \gamma_2^{\prime}, \gamma_1^{\prime}, \gamma_4^{\prime}\right).
\end{array}
\end{aligned}
\label{D4trans}
\end{equation}
Here, we omit the subscript $j$ for simplicity. For Majorana modes away from the symcenter, their subscripts $j$ should transform accordingly.

In this way, the Majorana pairs in Eq. \eqref{xyterm} are local mass terms of complex fermions. 
Therefore, the ground state of $H_0$ is clearly a topologically trivial product state. 
In fact, one can show that a free-fermion lattice model respecting the same set of lattice symmetries must always be topologically trivial~\cite{dihedral}.

Towards a nontrivial HOTSC, there should be intersite couplings centered at the maximal Wyckoff positions. 
Due to the crystalline 
symmetry constraint, the pairing between two Majoranas on the different sides of the reflection axis is forbidden since the reflection will inverse the direction of pairing. Hence, the intersite couplings need to be some four-fermion interaction.
Thus, we now explore the topological consequence of four-fermion interactions in our model.
The interacting Hamiltonian should be $D_4$ symmetric and trivially gapped in the periodic boundary conditions (PBC).
For the sake of convenience, at each maximal Wyckoff position $a$ or $c$, we redefine four complex fermion operators from Majorana operators ($k=1,2,3,4$),
\begin{align}
\begin{aligned}
&c_{k,ja}^\dag=\frac{1}{2}\left(\alpha_{k,j}+i\alpha_{k,j}'\right)\\
&c_{k,jc}^\dag=\frac{1}{2}\left(\gamma_{k,j}+i\gamma_{k,j}'\right).
\end{aligned}
\label{complex fermions}
\end{align}
This redefinition of complex fermions is equivalent to a change in basis or unitary transformation in the Hilbert space.
The particle number operators of these complex fermions are denoted as $n_{k,ja}=c_{k,ja}^\dag c_{k,ja}$ and $n_{k,jc}=c_{k,jc}^\dag c_{k,jc}$. 
We first introduce Hubbard interactions $H_U$ for these complex fermions, which are $D_4$-symmetric ($U>0$),
\begin{align}
H_{U}=U\sum_j\sum_{s=a,c}\sum_{k=1}^2\left(n_{k,js}-\frac{1}{2}\right)\left(n_{k+2,js}-\frac{1}{2}\right).
\label{Hubbard interaction}
\end{align}
In PBC, all Majorana operators are involved in $H_U$, and the occupations result in a fourfold ground state degeneracy (GSD) per $a/c$ site, i.e., $(n_{1,js},n_{3,js})$ or $(n_{2,js},n_{4,js})=(0,1)$ or $(1,0)$. 
This fourfold GSD can be effectively regarded as two pseudo-spin-$1/2$ degrees of freedom per site
\begin{equation}
\begin{array}{l}
\tau^{\mu}_{13, js}=\left(c_{1, js}^{\dagger}, c_{3, js}^{\dagger}\right) \sigma^{\mu}\left(\begin{array}{c}
c_{1, js} \\
c_{3, js}
\end{array}\right), \\
\tau^{\mu}_{24, js}=\left(c_{2, js}^{\dagger}, c_{4, js}^{\dagger}\right) \sigma^{\mu} \left(\begin{array}{c}
c_{2, js} \\
c_{4, js}
\end{array}\right),
\end{array}. \quad s=a,c.
\end{equation}
where $\sigma^{x}$, $\sigma^y$, and $\sigma^{z}$ are Pauli matrices.
Then, we introduce a spin-spin interaction at each maximal Wyckoff position ($J>0$),
\begin{align}
H_J=J\sum\limits_j\Big[\bs{\tau}_{13,ja}\ast\bs{\tau}_{24,ja}+\bs{\tau}_{13,jc}\ast\bs{\tau}_{24,jc}\Big].
\label{spin interaction}
\end{align}
Here, $\ast$ is defined as $\bs{S}_1\ast\bs{S}_2=S_1^xS_2^x+S_1^yS_2^z-S_1^zS_2^y$ to satisfy $D_4$ symmetry Eq. \eqref{D4trans}. 
$H_J$ lifts the GSD of $H_U$ to a nondegenerate ground state, which is a spin-singlet assembly. 
Note that when viewed on the previous basis (corresponding to real atomic sites), this ground state is not a product state because the alternative complex fermions defined in Eq. \eqref{complex fermions} are virtual and cannot be gapped by local mass terms.
At each maximal Wyckoff position $a$ or $c$, the nondegenerate ground state in the pseudospin basis is
\begin{align}
|\psi\rangle=\frac{1}{2}\Big(|\uparrow,\uparrow\rangle+i|\uparrow,\downarrow\rangle-i|\downarrow,\uparrow\rangle-|\downarrow,\downarrow\rangle\Big).
\end{align}
Likewise, each maximal Wyckoff position $a/c$ carries a linear representation. 
According to the LSM-type constraint, the interaction terms $H_U+H_J$ at each maximal Wyckoff position are suitable candidates for constructing a 2D HOTSC.

Following that, we investigate the topological properties of the proposed model in the topological nontrivial ($U/t \gg 1$) regime. 
We will see that Majorana zero modes appear at the geometric corners of the system in the open boundary condition (OBC) as a signature of the HOTSC phase.
As shown in Fig. \ref{FIG1}(b), for the lattice in OBC, all Majorana modes in the bulk are gapped out by $H_U + H_J$, whereas, Majorana modes on the boundary are not involved in interaction terms, resulting in gapless dangling boundary modes. 
Those 1D gapless boundary modes can be gapped out by a $D_4$-symmetric perturbation,
\begin{equation}
\begin{aligned}
H’= \epsilon\sum_j (&\alpha_{1,j}\alpha_{1,j}’\alpha_{2,j}\alpha_{2,j}’ + \alpha_{2,j}\alpha_{2,j}’\alpha_{3,j}\alpha_{3,j}’ 
\\+ &\alpha_{3,j}\alpha_{3,j}’\alpha_{4,j}\alpha_{4,j}’ + \alpha_{4,j}\alpha_{4,j}’\alpha_{1,j}\alpha_{1,j}’).
\end{aligned}
\end{equation}

\begin{figure*}
\includegraphics[width=0.18\textwidth]{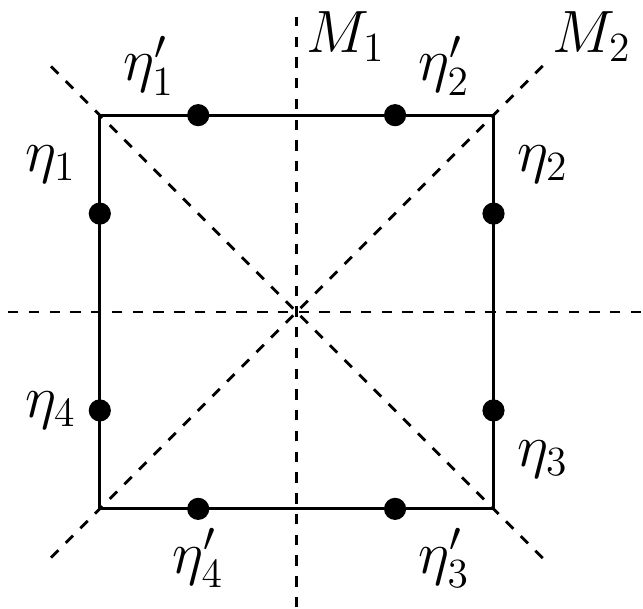}
\hspace{0.3cm}
\includegraphics[width=0.75\textwidth]{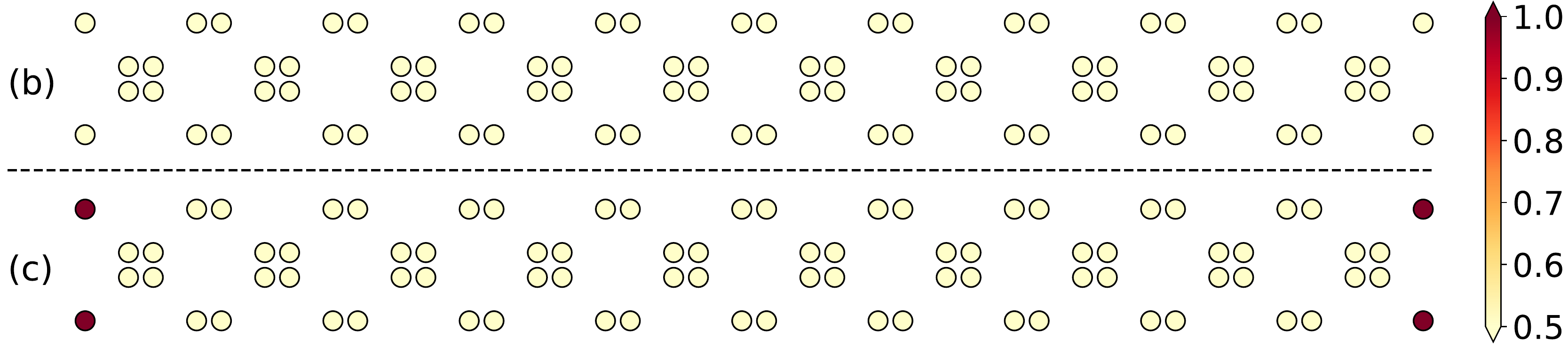}
\caption{
(a) Sketch of Majorana zero modes of $D_4$-symmetric HOTSC at geometric corners in OBC. 
(b) and (c) Evolution of Majorana corner modes of the Hamiltonian $H = H_0+H_U+H_J$ on a two-leg ladder when slightly tuned away from half-filling from (b) weakly interacting regime with $t = 1, U = J =0.1$ to (c) strongly interacting regime with $t = 1, U = J =10$. 
The color of each point indicates the ground-state density expectation value of complex fermions defined in Eq. \eqref{complex fermions}.
} 
\label{interaction}
\end{figure*}
The remaining zero-dimensional gapless boundary modes $\eta_k$ and $\eta_k'$ ($k=1,2,3,4$) are all strictly localized at the corner of the system [see Fig. \ref{interaction}(a)] with two at each corner and the symmetry properties,
\begin{align}
\begin{aligned}
&\bs{M}_1:\left\{\begin{aligned}
&\left(\eta_1,\eta_4\right)\leftrightarrow\left(\eta_2,\eta_3\right),\\
&\left(\eta_1',\eta_4'\right)\leftrightarrow\left(\eta_2',\eta_3'\right),
\end{aligned}\right.\\
&\bs{M}_2:\left(\eta_1,\eta_2,\eta_{3},\eta_{4}\right)\leftrightarrow\left(\eta_{3}',\eta_{2}',\eta_{1}',\eta_{4}'\right).
\end{aligned}
\label{D4 symmetry corner}
\end{align} 
These eight Majorana zero modes are protected by $D_4$ symmetry and cannot be gapped out by symmetric perturbations. 
The only possible way to gap out is to introduce a Majorana pair at each corner $i\eta_j\eta_j'$ ($j=1,2,3,4$).
However, these terms break the reflection operation in $D_4$ symmetry either diagonally or off diagonally [see dashed lines in Fig. \ref{interaction}(a)], e.g., $\bs{M}_2: i\eta_2\eta_2' \rightarrow -i\eta_2\eta_2'$.
As a result, all these Majorana corner modes are robust against $D_4$-symmetric perturbations, which feature a HOTSC phase.

We now numerically investigate the general case of our model with $t, U, J \neq 0$, and verify the Majorana corner modes of the HOTSC phase in the strongly interacting regime. 
Consider the interacting Hamiltonian $H = H_0+H_U+H_J$ on a $2\times20$ two-leg ladder geometry with OBC in the horizontal direction and PBC in the vertical direction. 
The computational basis is taken as the Fock basis with complex fermion operators defined as Eq. \eqref{complex fermions}. 
$H$ represents a half-filling system on this basis, and the electron density for the ground state is exactly $n_i=0.5$ at each site. 
Because $H$ is fully gapped in the bulk, perturbing $H$ with a small chemical potential $\mu$ allows us to reveal the zero modes.
When $\mu$ slightly deviates from 0, the variation of $n_i$ from half-filling reveals the density distribution of zero modes.
The density distribution with $\mu=0.2$ for both the weakly interacting regime and the strongly interacting regime is plotted in Fig. \ref{interaction}. 
In the weakly interacting regime [Fig. \ref{interaction}(b)], $\mu$ induces hardly any density variation, showing that there is no gapless edge mode and the phase is trivial. 
In the strongly interacting regime [Fig. \ref{interaction}(c)], however, an obvious density variation occurs, which is strictly localized at the corners. 
This demonstrates the existence of Majorana corner modes as well as the HOTSC phase.
The numerical result is consistent with our previous analysis in the free limit ($U,J \rightarrow 0$) and strong interaction limit ($t \rightarrow 0$).
The robustness of the Majorana corner modes against $D_4$-symmetric perturbations reveals the nontrivial topology of the model we have constructed. 
We emphasize that such a nontrivial topological property is a result of strong electron interactions.

\begin{figure}
\includegraphics[width=0.98\textwidth]{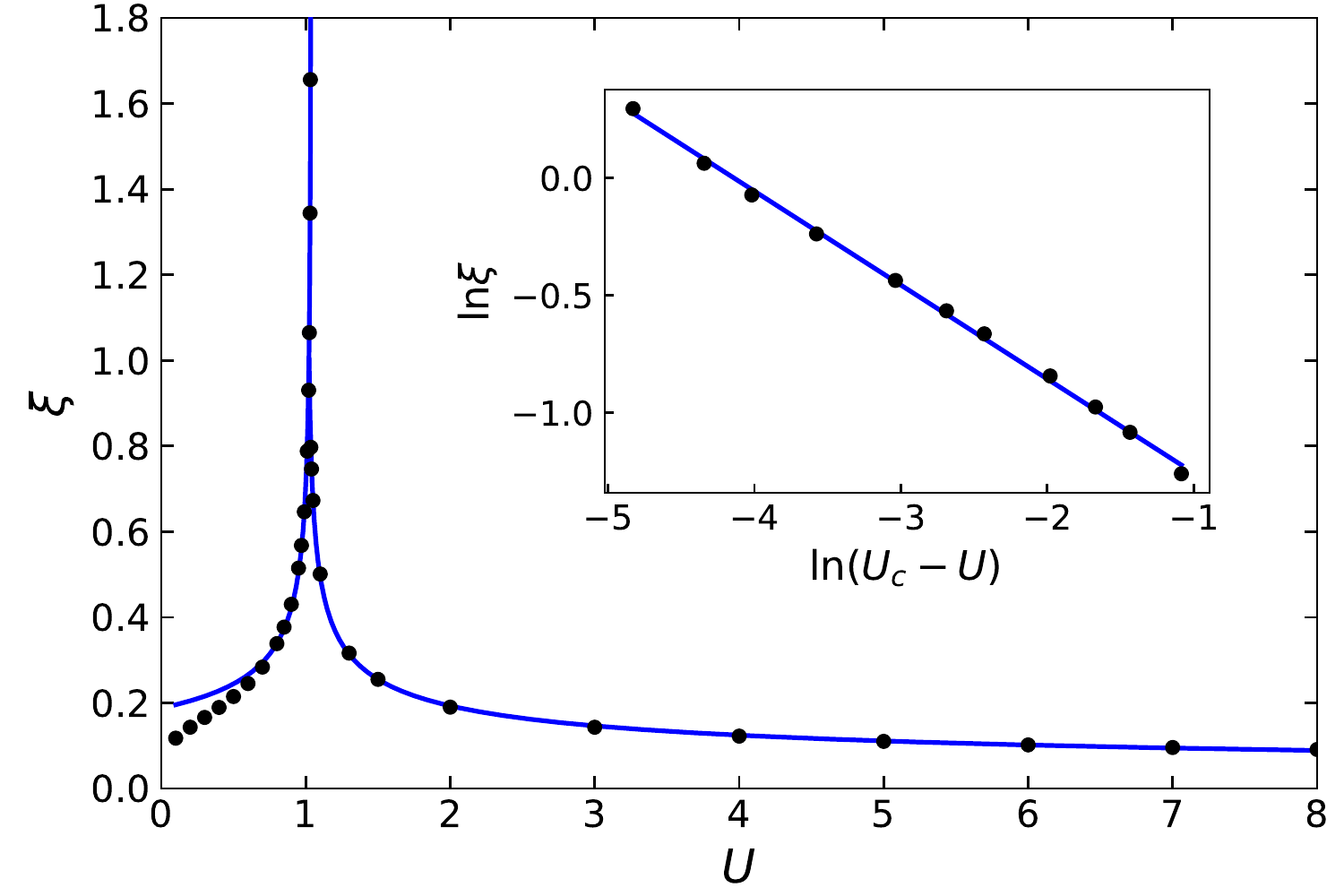}
\caption{
The correlation lengths $\xi$ of the matrix product state as a function of $U$ with fixed $t = 1$ and $J = U$. 
The power law divergence of $\xi$ near the critical point $U_c \sim 1.038$ signifies a quantum phase transition.
}
\label{correlation length}
\end{figure}

Furthermore, we find a topological quantum phase transition (QPT) by tuning the ratio of interaction strength to the intensity of Majorana pairs $U/t$. 
As previously stated, the system is a trivial product state when $U/t \ll 1$, whereas, it is a 2D $D_4$-symmetry-protected HOTSC when $U/t \gg 1$.
This QPT is characterized by the divergence of the correlation length $\xi$ on an infinite-length two-leg ladder, which is determined by the ratio between the two largest eigenvalues of the MPS transfer matrix \cite{Norbert2013}.
We fix $t = 1$ on a $2\times\infty$ lattice, tune the interaction strength $U$ ($J=U$), and calculate $\xi$ for various $U$'s ranging from $0$ to $8$ as shown in Fig. \ref{correlation length}.
$\xi$ diverges at $U_c \sim 1.038$, implying an unambiguous topological QPT between a trivial product state and an extended HOTSC. 
As shown in the inset of Fig. \ref{correlation length}, the correlation length exhibits a perfect power-law divergence for $U \lesssim U_c$, and we fit the critical exponent $\xi \propto |U - U_c|^{-0.40}$.
This QPT is also implied by the evolutions of Majorana corner modes for various $U/t$ ratios as shown in Fig. \ref{interaction}. 
According to the classification of 2D crystalline fermionic SPT phases in Ref. \cite{dihedral}, the lattice model we constructed in this Letter is the only possible 2D intrinsically interacting class-$D$ HOTSC, for both spinless and spin-$1/2$ fermions. 

\textit{Conclusion and discussion.} 
In this Letter, we construct an intrinsically interacting lattice model of 2D $D_4$-symmetric class-$D$ HOTSC using the LSM-type constraint.
An indispensable advantage of the LSM-type constraint is that it considerably simplifies the lattice model construction: only physical degrees of freedom forming linear representations of the total symmetry group at maximal Wyckoff positions are allowed. 
With the concrete lattice model, we study its Majorana corner modes, which are robust under symmetric perturbations, and find that there are two Majorana zero modes at each corner of the system. 
Subsequently, we perform MPS calculations on systems with two-leg ladder geometry in order to tackle the general interacting Hamiltonian. 
We see the stability of Majorana corner modes and find an unambiguous topological QPT between a 2D HOTSC and a trivial product state, controlled by the ratio between the intensity of interactions and Majorana pairs. 
The concrete lattice model we developed here is the only possible 2D intrinsically interacting class-$D$ HOTSC, and explicit and robust Majorana corner modes are experimentally relevant and can be directly measured with spectroscopic experiments on monolayer iron-selenide (FeSe, a strongly correlated material whose point-group symmetry is $D_4$).
Lattice model construction with the LSM-type constraint established in this Letter can be generalized to systems with any dimension and any spatial or internal symmetry for interacting fermionic systems.

\textit{Acknowledgements} -- 
Stimulating discussions with Z.Bi and M.Cheng are acknowledged. 
H.R.Z. and S.Y. are supported by the National Natural Science Foundation of China (NSFC) (Grants No. 12174214 and No. 92065205), the National Key R\&D Program of China (Grant No. 2018YFA0306504), and the Innovation Program for Quantum Science and Technology (Grant No. 2021ZD0302100).
J.H.Z. and Z.C.G. were supported by Direct Grant No. 4053462 from The Chinese University of Hong Kong and funding from Hong Kong’s Research Grants Council (Grant No. 14307621, ANR/RGC Joint Research Scheme No. A-CUHK402/18). 
R.X.Z. was supported by a startup fund at the University of Tennessee.

\bibliography{apssamp}

\clearpage

\appendix
\setcounter{equation}{0}
\newpage

\renewcommand{\thesection}{S-\arabic{section}} \renewcommand{\theequation}{S%
\arabic{equation}} \setcounter{equation}{0} \renewcommand{\thefigure}{S%
\arabic{figure}} \setcounter{figure}{0}

\onecolumngrid

\centerline{\large\textbf{Supplemental Materials of ``Intrinsically interacting higher-order topological superconductors''}}
\vskip0.8cm
\twocolumngrid
\maketitle

\section{\1. Total symmetry group in fermionic systems}
In this section, we show that the total symmetry group of a fermionic system is the central extension between the physical symmetry group $G_b$ and the fermion parity $\mathbb{Z}_2^f$. 
The parity of fermions in a fermionic system is always conserved, as the normal subgroup of the total symmetry of the system, $G_f$. 
The physical symmetry group $G_b$ is defined as the quotient group of $G_f$ and $\mathbb{Z}_2^f$: $G_b=G_f/\mathbb{Z}_2^f$. 
Inversely, the total symmetry group $G_f$ is the central extension of the physical symmetry group $G_b$ and fermion parity $\mathbb{Z}_2^f$
\begin{align}
0\rightarrow\mathbb{Z}_2^f\rightarrow G_f\rightarrow G_b\rightarrow0.
\end{align}
Different central extensions are characterized by different factor systems of this short exact sequence, which is labeled by a group 2-cocycle $\omega_2\in\mathcal{H}^2(G_b,\mathbb{Z}_2^f)$. 

\textbf{\underline{Lemma} (Factor System)} 
For a group $(G,\cdot)$, an Abelian group $(A,+)$, and a short exact sequence
\begin{equation}
0\rightarrow A\rightarrow X\rightarrow G\rightarrow0.
\label{AXG}
\end{equation}
There is a factor system of the short exact sequence Eq. (\ref{AXG}) that consists of the function $f$ and a homomorphism $\sigma$
\begin{align}
\left.
\begin{aligned}
f:G\times G~&\rightarrow~~~A,\\
(g,h)~&\mapsto f(g,h),
\end{aligned}
\right.~~~\left.
\begin{aligned}
\sigma:G~~&\rightarrow \text{End}(A),\\
g~~&\mapsto~~~\sigma_g,
\end{aligned}
\right.
\end{align}
so that the Cartesian product $G\times A$ becomes a group $X$ with multiplication $(g,a)*(h,b)=(g\cdot h,f(g,h)+a+\sigma_g(b))$.
End$(A)$ is the endomorphism of group $A$, and $f$ must be a group $2$-cocycle classified by $2$-group cohomology $f\in\mathcal{H}^2(G,A)$.

\section{\2. Ground-state degeneracy}
In this section, we show that the 1D Kitaev chain is incompatible with the $M^2=1$ reflection symmetry. 
We begin by cutting the Majorana chain in the middle (i.e. the reflection axis shown by the vertical dashed line), which results in two edge Majorana zero modes $\gamma_l$ and $\gamma_r$
\begin{align}
\includegraphics[width=0.8\linewidth]{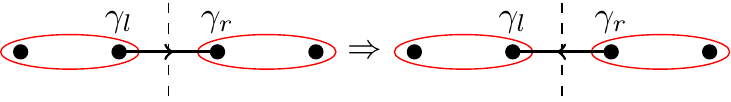}
\label{cut}
\end{align}
Here, the red ellipse and the black arrow represent the atomic site and the pair, respectively. 
Under reflection symmetry, the two Majorana zero modes transform into one another
\begin{align}
M:~\gamma_l\leftrightarrow\gamma_r.
\label{M^2=1}
\end{align}
So far, the full symmetry is preserved. 
We then attempt to glue the two half chains by removing the zero modes $\gamma_l$ and $\gamma_r$. 
However, we cannot glue them in a symmetric way because the only coupling term $i\gamma_l\gamma_r$ is odd under $M$. 
As a matter of fact, these two zero modes cannot be removed in a reflection symmetric way, even if additional 0D-block states are decorated. 
This is because the two dimensional Hilbert space spanned by $|0\rangle$ and $c^\dag|0\rangle$, with $c=(\gamma_l+i\gamma_r)/2$, forms a projective representation of the full symmetry group $\mathbb{Z}_2^f\times\mathbb{Z}_2$ at the reflection center (where $\mathbb{Z}_2^f=\{I,P_f\}$ represents fermion parity symmetry, and $\mathbb{Z}_2$ represents the reflection symmetry acting internally). 
Indeed, we have $\gamma_l=\sigma^x$, $\gamma_r=\sigma^y$, $P_f=\sigma^z$ in this Hilbert space, and
\begin{align}
\renewcommand\arraystretch{1.2}
M=\left(
\begin{array}{ccc}
0 & e^{-i\pi/4}\\
e^{i\pi/4} & 0
\end{array}
\right),
\end{align}
where $\sigma^i,~i=x,y,z$ are Pauli matrices. 
This representation fulfills the transformation Eq. (\ref{M^2=1}) and the condition $M^2=1$. 
We can easily verify that $MP_f=-P_fM$, i.e., a sufficient condition showing that the local Hilbert space is a projective representation of the symmetry group $\mathbb{Z}_2^f\times\mathbb{Z}_2$. 
Hence, the two-fold ground-state degeneracy of $|0\rangle$ and $c^\dag|0\rangle$ cannot be lifted, even if additional 0D-block states (i.e., linear representations) are attached.
Accordingly, the 1D Majorana chain is not compatible with the $M^2=1$ reflection symmetry.

Next, we cut the Majorana chain according to Eq. (\ref{cut}) for $M^2=-1$ reflection symmetry, and the two Majorana zero modes have the following reflection symmetry properties
\begin{align}
M:~\gamma_l\mapsto\gamma_r,~~\gamma_r\mapsto-\gamma_l.
\end{align}
Now the coupling term $i\gamma_l\gamma_r$ is even under $M$. 
Therefore, the 1D Kitaev chain is compatible with the $M^2=-1$ reflection symmetry.
The spin of fermions plays a central role in characterizing the crystalline topological phases.

\section{\3.~LSM-type constraint for topological crystalline phases}
In the main text, we build the lattice model of the 2D $D_4$-symmetric class-$D$ higher-order TSC using the LSM-type constraint for topological crystalline phases. 
In this section, we demonstrate the LSM-type constraint by a 2D $D_2$-symmetric lattice with spinless fermions. 

The LSM-type constraint for topological crystalline phases in 2D interacting fermionic systems is defined as follows: for a $SG$-symmetric lattice ($SG$ is the group of a 2D crystalline symmetry), a gapped, non-degenerate ground state requires that the system can be adiabatically connected to a state with an integer multiple of linear representations of the total symmetry group at maximal Wyckoff positions per unit cell. 

We shall consider different $D_2$-symmetric lattice constructions, and show that they can be gapped out trivially if and only if the physical degrees of freedom at each maximal Wyckoff position form a linear representation of the total symmetry group. 

\begin{figure}[tbp]
\includegraphics[width=0.7\textwidth]{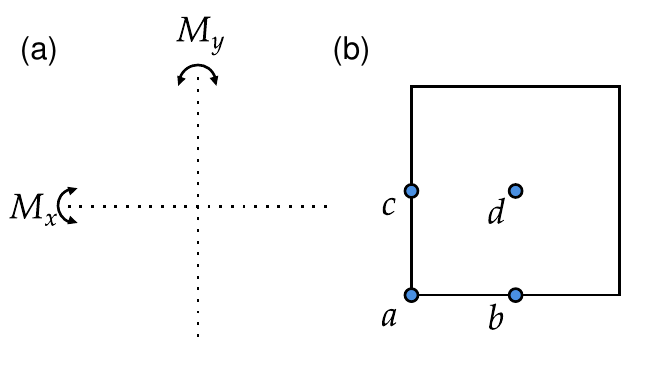}
\caption{
(a) Reflection generators of $D_2$ symmetry $\bs{M}_x$ and $\bs{M}_y$. 
(b) Maximal Wyckoff positions of $D_2$ symmetry $a,b,c,d$. 
There is a $D_2$ symmetry on each of them.
}
\label{Wyckoff}
\end{figure}

We first consider two Majorana zero modes per unit cell. 
Without loss of generality, they are all assigned to the maximal Wyckoff position denoted by $a$ (see Fig. \ref{Wyckoff}). 
There are several possibilities for assigning these Majorana zero modes.

1. Both $\gamma^A$ and $\gamma^B$ are assigned at the center of $D_2$. 
In this case, the $D_2$ symmetry properties of $\gamma^A$ and $\gamma^B$ in each unit cell are trivial.
\begin{align}
\bs{M}_x,\bs{M}_y:~~\left(\gamma^A,\gamma^B\right)\mapsto\left(\gamma^A,\gamma^B\right).
\end{align}
It is easy to verify that the following Hamiltonian is symmetric under $D_2$ symmetry with a fully-gapped spectrum for both the open boundary condition (OBC) and the periodic boundary condition (PBC)
\begin{align}
H=i\sum\limits_j\gamma_j^A\gamma_j^B.
\end{align}
Meanwhile, it is straightforward to see that $\gamma^A$ and $\gamma^B$ in each unit cell form a (trivial) linear representation of the total symmetry group $\mathbb{Z}_2^f\times D_2$. 
It is consistent with the LSM-type constraint.

\begin{figure}[tbp]
\includegraphics[width=0.99\textwidth]{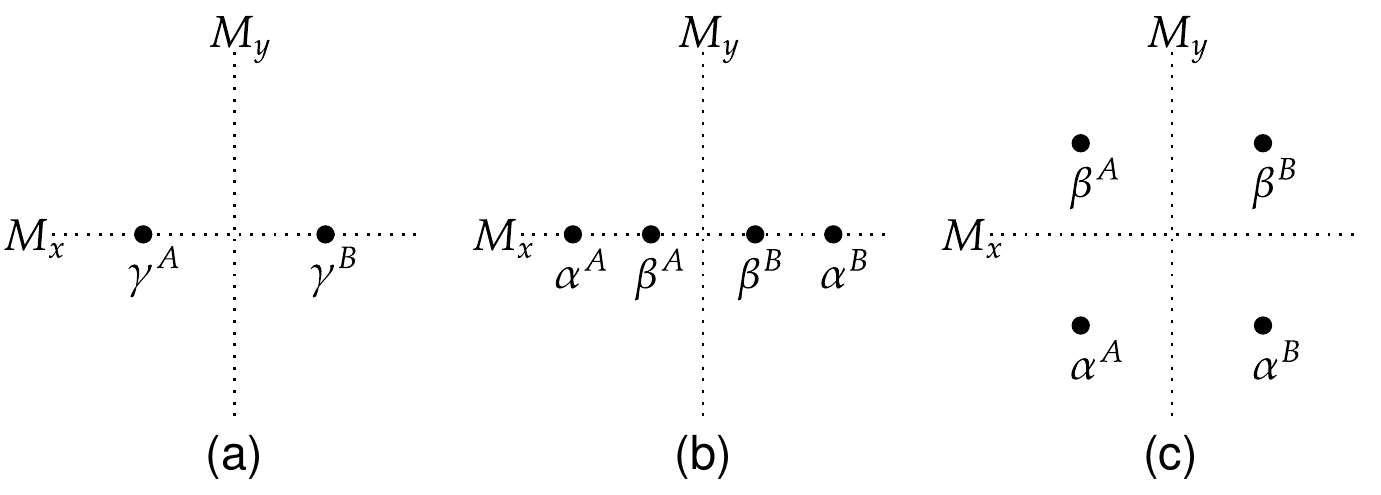}
\caption{
Different ways of assigning Majorana zero modes near maximal Wyckoff positions. 
They form different representations of the total symmetry group $\mathbb{Z}_2^f \times D_2$. 
(a, c) A projective representation. 
(b) A linear representation.
}
\label{assignment}
\end{figure}

2. $\gamma^A$ and $\gamma^B$ are assigned on the reflection axis (e.g., $\bs{M}_x$-axis), as shown in Fig. \ref{assignment}(a). 
In this case, the $D_2$ symmetry properties of $\gamma^A$ and $\gamma^B$ in each unit cell are
\begin{align}
\begin{aligned}
&\bs{M}_x:~\left(\gamma^A,\gamma^B\right)\mapsto\left(\gamma^A,\gamma^B\right),\\
&\bs{M}_y:~\left(\gamma^A,\gamma^B\right)\mapsto\left(\gamma^B,\gamma^A\right).
\end{aligned}
\end{align}
To gap out the system, one may try adding pairing terms like $i\gamma_j^A\gamma_{j+\hat{x}}^B$, $i\gamma_j^B\gamma_{j+\hat{x}}^A$, or $i\gamma_j^A\gamma_j^B$, but these terms are odd under $\bs{M}_y$ and thus forbidden by symmetry. 
An allowed symmetric Hamiltonian is
\begin{align}
H =i\sum\limits_j\left(\gamma_j^A\gamma_{j+\hat{x}}^A-\gamma_j^B\gamma_{j+\hat{x}}^B\right).
\end{align}
Nevertheless, the spectrum of this Hamiltonian is gapless for both PBC and OBC. 
Thus, we fail to gap out the system symmetrically. 
On the other hand, $\gamma^A$ and $\gamma^B$ form a projective representation of $\mathbb{Z}_2^f\times D_2$ in each unit cell. 
More precisely, they form a projective representation of $\mathbb{Z}_2^f\times\mathbb{Z}_2=\mathbb{Z}_2^f\times\{I,\bs{M}_y\}$ ($I$ is identity) as a subgroup of $\mathbb{Z}_2^f\times D_2$. 
Then the LSM-type constraint indicates that it is impossible to construct a Hamiltonian with a unique gapped symmetric ground state.

Next, in Fig. \ref{assignment}(b) we consider two copies of the aforementioned Majorana fermions $\alpha^A$, $\alpha^B$, $\beta^A$, and $\beta^B$ per unit cell with $D_2$ symmetry properties
\begin{align}
\begin{aligned}
&\bs{M}_x:~\left(\alpha^A,\alpha^B,\beta^A,\beta^B\right)\mapsto\left(\alpha^A,\alpha^B,\beta^A,\beta^B\right),\\
&\bs{M}_y:~\left(\alpha^A,\alpha^B,\beta^A,\beta^B\right)\mapsto\left(\alpha^B,\alpha^A,\beta^B,\beta^A\right).
\end{aligned}
\end{align}
The following Hamiltonian satisfies all symmetries and has a fully gapped spectrum for both PBC and OBC
\begin{align}
H=i\sum\limits_j\left(\alpha_j^A\beta_j^A+\alpha_j^B\beta_j^B\right).
\end{align}
This Hamiltonian has a gapped, non-degenerate ground state. 
Meanwhile, $\left(\alpha^A,\alpha^B\right)$ and $\left(\beta^A,\beta^B\right)$ are both projective representations of $\mathbb{Z}_2^f\times\mathbb{Z}_2$. 
The projective representations of $\mathbb{Z}_2^f\times\mathbb{Z}_2$ are classified by group 2-cohomology
\begin{align}
\mathcal{H}^2\left[\mathbb{Z}_2^f\times\mathbb{Z}_2,U(1)\right]=\mathbb{Z}_2,
\end{align}
i.e., there is only one nontrivial projective representation of $\mathbb{Z}_2^f\times\mathbb{Z}_2$. 
As a result, $\left(\alpha^A,\alpha^B,\beta^A,\beta^B\right)$ forms a linear representation of $\mathbb{Z}_2^f\times\mathbb{Z}_2$ (and $\mathbb{Z}_2^f\times D_2$), so it is possible for us to construct a Hamiltonian with a unique gapped symmetric ground state.

Furthermore, in Fig. \ref{assignment} (c) we consider $\alpha^A$, $\alpha^B$, $\beta^A$, and $\beta^B$ assigned away from both reflection axes with $D_2$ symmetry properties
\begin{align}
\begin{aligned}
&\bs{M}_x:~\left(\alpha^A,\alpha^B,\beta^A,\beta^B\right)\mapsto\left(\beta^A,\beta^B,\alpha^A,\alpha^B\right),\\
&\bs{M}_y:~\left(\alpha^A,\alpha^B,\beta^A,\beta^B\right)\mapsto\left(\alpha^B,\alpha^A,\beta^B,\beta^A\right).
\end{aligned}
\end{align}
Again, one may try to add pairing terms like $H_x$ or $H_{xy}$ to gap out the system
\begin{align}
\begin{aligned}
&H_x=i\sum\limits_j\left(\beta_j^B\alpha_{j+\hat{x}}^A-\beta_j^A\alpha_{j+\hat{x}}^B\right),\\
&H_{xy}=i\sum\limits_j\left(\beta_j^B\alpha_{j+\hat{x}+\hat{y}}^B+\alpha_j^A\beta_{j-\hat{x}+\hat{y}}^A\right).
\end{aligned}
\end{align}
However, both $H_x$ and $H_{xy}$ break $D_2$ symmetry explicitly.
The minimal form of a symmetric Hamiltonian is
\begin{align}
H=i\sum\limits_{j}\left(\alpha_j^A\alpha_{j+\hat{x}}^A-\alpha_j^B\alpha_{j+\hat{x}}^B+\beta_j^A\beta_{j+\hat{x}}^A-\beta_j^B\beta_{j+\hat{x}}^B\right).\nonumber
\end{align}
Nevertheless, this Hamiltonian has a gapless spectrum for both PBC and OBC.

We also show that $\alpha^A$, $\alpha^B$, $\beta^A$, and $\beta^B$ per unit cell form a projective representation of $D_2$, i.e., a spin-1/2 degree of freedom. 
This is because as a discrete subgroup of $SO(3)$, the spin-1/2 degree of freedom is the only nontrivial projective representation of $D_2$. We now define two complex fermions
\begin{align}
c_1^\dag=\frac{1}{2}\left(\alpha^A+i\alpha^B\right),~~c_2^\dag=\frac{1}{2}\left(\beta^A+i\beta^B\right).
\end{align}
with the following $D_2$ symmetry properties
\begin{align}
\begin{aligned}
&\bs{M}_x:~\left(c_1^\dag,c_2^\dag\right)\mapsto\left(c_2^\dag,c_1^\dag\right),\\
&\bs{M}_y:~\left(c_1^\dag,c_2^\dag\right)\mapsto\left(ic_1,ic_2\right).
\end{aligned}
\end{align}
We introduce a Hubbard interaction ($U>0$ and $n_j=c_j^\dag c_j$) to gap out all these fermions
\begin{align}
H_U=U\left(n_1-\frac{1}{2}\right)\left(n_2-\frac{1}{2}\right).
\end{align}
$c_1^\dag|0\rangle$ and $c_2^\dag|0\rangle$ are two degenerate ground states of $H_U$ that can be treated as a spin-1/2 degree of freedom. 
We introduce
\begin{align}
\tau^\mu=\left(c_1^\dag,c_2^\dag\right)\sigma^\mu\left(
\begin{array}{ccc}
c_1\\
c_2
\end{array}
\right).
\end{align}
with the $D_2$ symmetry properties
\begin{align}
\begin{aligned}
&\bs{M}_x:~\left(\tau^x,\tau^y,\tau^z\right)\mapsto\left(\tau^x,-\tau^y,-\tau^z\right),\\
&\bs{M}_y:~\left(\tau^x,\tau^y,\tau^z\right)\mapsto\left(-\tau^x,\tau^y,-\tau^z\right).
\end{aligned}
\end{align}

\begin{figure}[tbp]
\includegraphics[width=0.9\textwidth]{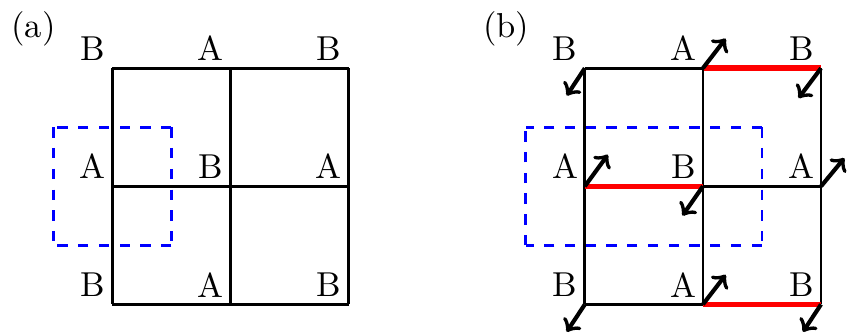}
\caption{
Symmetry breaking ground state of the model with one spin-1/2 degree of freedom per site on a square lattice. 
(a) Without translation symmetry breaking, there is one spin-1/2 degree of freedom per unit cell (as elucidated by the blue dashed rectangle). 
(b) With translation symmetry broken, the unit cell is doubled. Each thick red link represents a spin-singlet pair.
}
\label{VBS}
\end{figure}

We can see that $D_2$ symmetry prohibits all possible mass terms (like $m\tau^z$). 
As a result, two spin states $|\uparrow\rangle$ and $|\downarrow\rangle$ of $\bs{\tau}$ are degenerate. 
It indicates that the four Majorana fermions $\alpha^A$, $\alpha^B$, $\beta^A$, and $\beta^B$ form a projective representation of the $D_2$ symmetry group (and thus a projective representation of $D_2\times\mathbb{Z}_2^f$). 
The only way to lift this GSD is to break the translation symmetry, which is accomplished by dividing a bipartite lattice into two sub-lattices $A$ and $B$, as shown in Fig. \ref{VBS}(b). 
Without breaking translation symmetry ($A/B$ sub-lattices are equivalent, see Fig. \ref{VBS}(a)), there will be a $2$-fold GSD at each site, resulting in a flat band with the lowest energy in the thermodynamic limit. 
Therefore, we introduce $A/B$ sub-lattices and a spin exchange interaction in each unit cell
\begin{align}
H_J=J\sum\limits_j\bs{\tau}_j^A\cdot\bs{\tau}_j^B,~~J>0.
\end{align}
It is obvious that $H_J$ can lift all ground state degeneracy, resulting in a non-degenerate effective valence-bond solid (VBS) ground state with a spin-singlet pair in each unit cell [see Fig. \ref{VBS}(b)]. 
With translation symmetry broken, there are two spin-$1/2$ degrees of freedom in each doubled unit cell as two projective representations of the $D_2$ group. 
They together form a linear representation of $\mathbb{Z}_2^f \times D_2$ per unit cell. 
This is consistent with the LSM-type constraint.

In addition, if each unit cell contains two copies of the aforementioned four Majorana fermions $(\alpha^A,\alpha^B,\beta^A,\beta^B)$ and $(\gamma^A,\gamma^B,\xi^A,\xi^B)$, we expect to be able to gap out the system symmetrically because they form a linear representation per unit cell. 
In this case, the $D_2$ symmetry properties of these Majorana fermions are
\begin{align}
\begin{aligned}
&\bs{M}_x:~\left\{
\begin{aligned}
&\left(\alpha^A,\alpha^B,\beta^A,\beta^B\right)\mapsto\left(\alpha^B,\alpha^A,\beta^B,\beta^A\right),\\
&\left(\gamma^A,\gamma^B,\xi^A,\xi^B\right)\mapsto\left(\gamma^B,\gamma^A,\xi^B,\xi^A\right),
\end{aligned}
\right.\\
&\bs{M}_y:~\left\{
\begin{aligned}
&\left(\alpha^A,\alpha^B,\beta^A,\beta^B\right)\mapsto\left(\beta^B,\beta^A,\alpha^B,\alpha^A\right),\\
&\left(\gamma^A,\gamma^B,\xi^A,\xi^B\right)\mapsto\left(\xi^B,\xi^A,\gamma^B,\gamma^A\right).
\end{aligned}
\right.
\end{aligned}
\end{align}
We can construct a symmetric Hamiltonian with minimal terms $H_{xy}$ (diagonal and off-diagonal terms), $H_x$ (horizontal terms) and $H_y$ (vertical terms)
\begin{align}
\begin{aligned}
&H_{xy}=i\sum\limits_{j}\xi_j^B\alpha_{j+\hat{x}+\hat{y}}^B-\alpha_j^A\xi_{j-\hat{x}+\hat{y}}^A\\
&\quad \quad \quad \quad
+\gamma_j^B\beta_{j-\hat{x}-\hat{y}}^B+\gamma_j^A\beta_{j-\hat{x}+\hat{y}}^A ,\\
&H_x=i\sum\limits_j\beta_j^B\gamma_{j+\hat{x}}^A-\xi_j^A\alpha_{j+\hat{x}}^B+\beta_j^A\gamma_{j+\hat{x}}^B-\xi_j^B\alpha_{j+\hat{x}}^A ,\\
&H_y=i\sum\limits_j\beta_j^B\xi_{j+\hat{y}}^A-\gamma_j^A\alpha_{j+\hat{y}}^B+\alpha_j^A\gamma_{j+\hat{y}}^B-\xi_j^B\beta_{j+\hat{y}}^A .
\end{aligned}
\end{align}
The total Hamiltonian $H=H_x+H_y+H_{xy}$ has a fully gapped spectrum under PBC. 
Therefore, for 2D $D_2$-symmetric systems with spinless fermions, a gapped, non-degenerate ground state requires that the system can be adiabatically connected to a state with an integer multiple of linear representations of $\mathbb{Z}_2^f\times D_2$ group at maximal Wyckoff positions per unit cell.
The above procedure can be generalized to all other crystalline symmetries, for both spinless and spin-1/2 fermions.

\end{document}